\documentclass[aps,reprint,showpacs,showkeys,superscriptaddress]{revtex4-1} 
\usepackage[utf8]{inputenc}
\usepackage{cmap} 
\usepackage[T1]{fontenc}
\usepackage{microtype}
\usepackage{amsmath,amssymb,mathtools}
\usepackage{txfonts}

\usepackage{mathrsfs}
\usepackage{graphicx,grffile,textcomp}
\graphicspath{{figure/}}

\usepackage{textcomp}
\usepackage{booktabs,tabularx,dcolumn}
\newcolumntype{d}[1]{D{.}{.}{#1}}

\usepackage{url,hyperref}
\usepackage[usenames,dvipsnames]{xcolor}
\hypersetup{colorlinks=true, linkcolor=BrickRed, urlcolor=blue!50!black, citecolor=blue!50!black}

\usepackage[capitalize]{cleveref}
\crefrangelabelformat{equation}{(#3#1#4)$-$(#5#2#6)}

\usepackage{placeins}

\renewcommand\leq{\leqslant}

\renewcommand\rho\varrho
\renewcommand\vec[1]{\textrm{\bfseries #1}}

\begin{document}
\title{Phase separation around heated colloid in bulk and under confinement}

\author{Sutapa Roy}
\email{sutapa@is.mpg.de}
\affiliation{Max-Planck-Institut f\"{u}r Intelligente Systeme,
   Heisenbergstr.\ 3,
   70569 Stuttgart,
   Germany}
\affiliation{IV. Institut f\"{u}r Theoretische Physik,
   Universität Stuttgart,
   Pfaffenwaldring 57,
   70569 Stuttgart,
   Germany}
   
\author{Anna Macio\l ek}
\email{amaciolek@ichf.edu.pl}
\affiliation{Institute of Physical Chemistry, Polish Academy of Sciences, 
   Kasprzaka 44/52, PL-01-224 Warsaw, Poland}  

\date{\today}

\begin{abstract}
We study the non-equilibrium coarsening dynamics of a binary liquid solvent around 
a colloidal particle in a presence of a time-dependent temperature gradient that 
emerges after temperature quench of a suitably coated colloid surface. The solvent 
is maintained at its critical concentration and the colloid is fixed in space. The 
coarsening patterns near the surface are shown to be strongly dependent on the colloid 
surface adsorption properties and on the temperature evolution. The temperature 
gradient alters the morphology of a binary solvent near the surface of a colloid as 
compared to the coarsening proceeding at constant temperature everywhere. We also 
present results for the evolution of coarsening in thin films with confining surfaces 
preferring one species of the binary liquid mixture over the other. Confinement leads 
to a faster phase segregation process and formation of a bridge connecting the colloid 
and both the confining walls. 
\end{abstract}

\maketitle

\section{Introduction}\label{introduction}

Phase separation of binary mixtures induced by a temperature quench into the miscibility 
gap is a subject of continuous research activity driven by application perspectives, e.g., 
the buildup of nanostructured materials of well defined structure~\cite{AY,Torino:2016}. 
In the latter context, the effects of surface and confinement on a coarsening process are 
of particular interest as they may be used to control the structure formation. If the 
surface has a preference for one of the two components of a binary mixture, the fluid 
structures emerging after the homogeneous temperature quench at a critical concentration 
are essentially distinct from those due to spinodal decomposition \cite{binder2006,puri2005,BPDH}. 
This is because the adsorption of the preferred component at the surface affects the 
fluctuations of concentration away from the surface. In the presence of confinement, e.g., 
in thin films, the coarsening process becomes complex due to the interplay between finite-size 
and surface adsorption effects ~\cite{BPDH}. Addition of colloidal particles to a binary 
liquid undergoing demixing via spinodal decomposition widens the possibility for controlling 
pattern formation. It has been demonstrated that colloidal particles significantly curtail 
coarsening \cite{Chung,Herzig} and in the case of adsorptionwise neutral colloids can lead 
to formation of interesting soft-solid materials, called 'bijels' \cite{Herzig,Stratford}. 
By using Janus colloids, with a difference in adsorption preference between its two hemispheres, 
one can create regular lamellar structures \cite{Krekhov:2013,Iwashita:2013}.

Application of {\it local} temperature quenches instead of spatially homogeneous ones creates 
temperature gradients, which strongly couple to the local composition of a binary mixture and 
alter the mechanism of coarsening. For example, if such a quench propagates through space 
over the time it may lead to macroscopic coarsening with patterns that are different from 
those seen in spatially homogeneous quenching \cite{Kurita:2017}. Local temperature quenches 
can be realized by direct laser heating \cite{Kurita:2017,Koehler} of parts of the binary 
liquid mixtures or by optical heating \cite{Koehler,Sano, wurger2013} of the surface of the suitably 
coated colloid suspended in such mixtures. One might think of using optically heated colloids 
for the buildup of soft solids, therefore, it is interesting to know how local is the phase 
separation around each colloid and how the coarsening process depends on the adsorption 
preference of the surface of a colloid. Here we address these problems by using numerical 
simulations. We focus on the early-stage of a coarsening process in which diffusive dynamics 
dominates. We are also interested in the coarsening mechanism around the colloids which are not 
kept in bulk, but are confined between surfaces exhibiting an adsorption preference for one of 
the two components of a binary solvent. Such studies are relevant for typical experimental 
realizations in which the samples cells are in a slab geometry. The effects of such confinement 
on the early-time non-equilibrium process of concentration gradient formation around the 
colloid are difficult to foresee. What makes it particular complicated is the presence of 
the temperature field coupled to the local concentration field.

Local quenches of a binary mixture as realized by the optical heating of the surface of the 
suspended, suitably coated colloid bear particular relevance for experimental studies of 
moving Janus colloids  \cite{bechinger2011,buttinoni2012,ruben2016,solano2017}. However, for 
these systems the transient dynamics of the binary solvent at early times is hardly studied 
\cite{roy2017}. In Ref.~\cite{roy2017}, which we coauthored, both Janus and homogeneous 
spherical particles have been considered for off-critical compositions of a binary solvent. 
In the present paper we focus on the homogeneous colloid and the critical composition of the 
solvent, which are relevant for fabrication of bicontinuous structures such as bigels. In order 
to mimic different physical situations, we consider two types of boundary conditions for the 
temperature field at the boundary of the simulation box  and compare the corresponding coarsening 
patterns.We also extend the model studied in Ref.~\cite{roy2017}  in order to assess the role of 
the Soret effect for pattern formation. In Ref.~\cite{Koehler} it has been argued that taking 
into account this effect is necessary to reproduce in numerical simulations the essential 
spatial and temporal phenomena occurring after local heating a polymer blend in the two phase 
region. 

Our paper is organized as follows: in Sec.\ref{model} we explain the phenomenological model 
considered and the numerical techniques used in our work. Section \ref{results} presents our 
analytical results for the time-dependent temperature profiles for different boundary conditions. 
There, we also explain our numerical findings on the non-equilibrium dynamics of structure 
formation around a suspended colloidal particle under a temperature gradient. All results 
correspond to the critical concentration of the binary solvent and thermal quench inside the 
binodal region. Section \ref{results} also contains results for phase separation around a 
colloid kept in confinement in a slab geometry. There, we  provide a brief discussion on the 
influence of Soret effect on the structure formation process. Finally Sec IV. summarizes the 
paper with a brief perspective of future works.

\section{Model}\label{model}

A local temperature quench of the surface of a colloidal particle gives rise to a temperature 
front, which propagates away from the surface. This can be described by time-dependent temperature 
field $T(\vec r,t)$. Temperature field couples strongly to the local concentration of a binary 
solvent. We introduce the order parameter (OP) field $\psi(\vec r,t)$ as a difference 
$\psi(\vec r,t)=c_A(\vec r,t)-c_B(\vec r,t)$ in the concentration of components $A$ and $B$ of 
a binary mixture, where $c_{A,B}=\rho_{A,B}/(\rho_{A}+\rho_{A,B})=N_{A,B}/N$ and 
$\rho_{A,B}=N_{A,B}/V$ are the local number densities. This OP is conjugated to the chemical 
potential difference $\mu=\mu_A-\mu_B$ between the chemical potentials of species $A$ and $B$. 
For phase separation driven by diffusion, we assume that the time $t$ evolution of both fields 
is governed by the Cahn-Hilliard-Cook (CHC) equation or Model B \cite{Hohenberg1977} based on 
the Ginzburg-Landau free energy functional, in conjunction with the heat diffusion equation. 
The Ginzburg-Landau free energy functional of the solvent OP is given by
\begin{equation}\label{CHC1} 
  \frac{\mathcal{F}}{k_BT_c} = \int \frac{d^{d}r}{\upsilon}\Bigl[\frac{1}{2}C\bigl(\nabla \psi(\vec r)\bigr)^{2} + \frac{1}{2}{a} \psi(\vec r)^{2}
  + \frac{1}{4} u  \psi(\vec r)^{4}\Bigr],
\end{equation}
with $a \propto (T-T_c)/T_c$ for an upper critical point, which we consider here. $T_c$ is 
the demixing critical temperature of the binary liquid mixture. $\upsilon$ is a microscopic 
volume unit which is typically taken to be equal to $a_0^3$ where $a_0$ is a microscopic length 
scale, e.g., size of molecules of the solvent. For a spatially varying temperature field we 
replace the parameter $a$ in Eq.~(\ref{CHC1}) by $\tilde T(\vec r)={\mathcal A} (T(\vec r)-T_c)/T_c)$, 
where $\mathcal A$ is a constant number. Then using the continuity equation for the evolution 
of the OP, which is conserved, together with a generalized Fick's law 
\begin{eqnarray}\label{CHC2} 
  &&\frac{\partial \psi(\vec r,t)}{\partial t} = -{\boldsymbol{\nabla}} \cdot {\boldsymbol j}({\boldsymbol r},{t}) \nonumber \\ 
  &=& - {\boldsymbol{\nabla}} \cdot  [{-M {\boldsymbol \nabla}\mu({\boldsymbol r},{t})}] =  
  M {\boldsymbol \nabla^2} \frac{\delta {\mathcal{F}}[{\psi}]}{\delta {\psi}({\boldsymbol r},{t})}
\end{eqnarray}
and adding a Gaussian random thermal noise $\eta(\vec r,t)$, one obtains the modified CHC 
equation with a spatio-temporal temperature field $\tilde T(\vec r,t)$ ~\cite{essery1990}
\begin{eqnarray}\label{CHC3} 
  \frac{\partial \psi(\vec r,t)}{\partial t} &=& \frac{M}{\upsilon} k_B T_c \nabla ^2 \Big ({\tilde T}(\vec r,t) \psi(\vec r,t) + 
  u\psi^3(\vec r,t) - C\nabla^2 \psi(\vec r,t) \Big) \nonumber \\ 
  &+& \eta(\vec r,t).
\end{eqnarray}
$M>0$ is the diffusive mobility which is assumed to be spatially constant and independent 
of the local concentration. Assuming local equilibrium, the noise $\eta(\vec r,t)$ conserving 
the OP obeys the following fluctuation-dissipation relation 
\begin{equation}\label{CHC4} 
  \langle \mu(\vec r,t) ~\mu(\vec r', t')\rangle= -2(M/\upsilon) k_B T(\vec r) \nabla^2 \delta (\vec r-\vec r') \delta (t -t').
\end{equation}
For the system at homogeneous temperature $T$, the correlation length (below $T_c$) and the 
characteristic time are given by $\xi_{-}=\sqrt{C/2|\tilde T|}$ and $t_0= 2\xi^2_-(T)/(D_m(T)|\tilde T|)$, 
respectively, where $D_m(T)=(M/\upsilon) k_B T$ is the interdiffusion constant of a binary solvent. 
We normalize the length, time and $\psi$ by $\xi_{-}$, $t_0$, and $\psi_1 = \sqrt{|\tilde T_1| /u}$, 
where $\psi_1$ corresponds to the concentration after phase separation, taken {\it at the quench 
temperature} $T=T_1<T_c$. We then obtain the dimensionless form of Eq.~(\ref{CHC3})
\begin{equation}\label{CHC6} 
  \frac{\partial \psi(\vec r,t)}{\partial t} = \nabla ^2 \Big (\frac{{\tilde T}(\vec r,t)}{\tilde T_1} \psi(\vec r,t) + 
  \psi^3(\vec r,t) - \nabla^2 \psi(\vec r,t) \Big)+ \eta(\vec r,t),
\end{equation}
where the thermal noise $\eta(\vec r,t)$ is expressed in units of  $\eta_0 = \psi_1/t_0$. 
After the colloid surface is cooled, because of heat flow the surrounding solvent temperature 
also changes with time. Time evolution of the temperature field is dictated by the heat diffusion 
equation
\begin{equation}\label{CHC7} 
  \frac{\partial \tilde T(\vec r,t)}{\partial t} = {\mathcal D} \nabla ^2 \tilde T(\vec r,t),
\end{equation}
where, $\mathcal D=D_{th}/(|\tilde T_1| D_m)$ is the so-called Lewis number and $D_{th}$ is 
the thermal diffusivity of the solvent. Equations (\ref{CHC6}) and (\ref{CHC7}) were first 
considered in Ref.~\cite{essery1990} to study phase separation near a planar surface subjected 
to the temperature quench. The authors of Ref.~\cite{binder2013} derived a CHC equation starting 
from a master equation for a Kawasaki spin-exchange kinetic Ising model \cite{Kawasaki} with a 
space-dependent temperature. They have obtained a more complex expression with additional terms 
involving coupling between ${\boldsymbol \nabla}T$ and ${\boldsymbol \nabla\psi}$ and higher order 
temperature derivatives. 

\begin{figure}
\centering
\includegraphics*[width=0.5\textwidth]{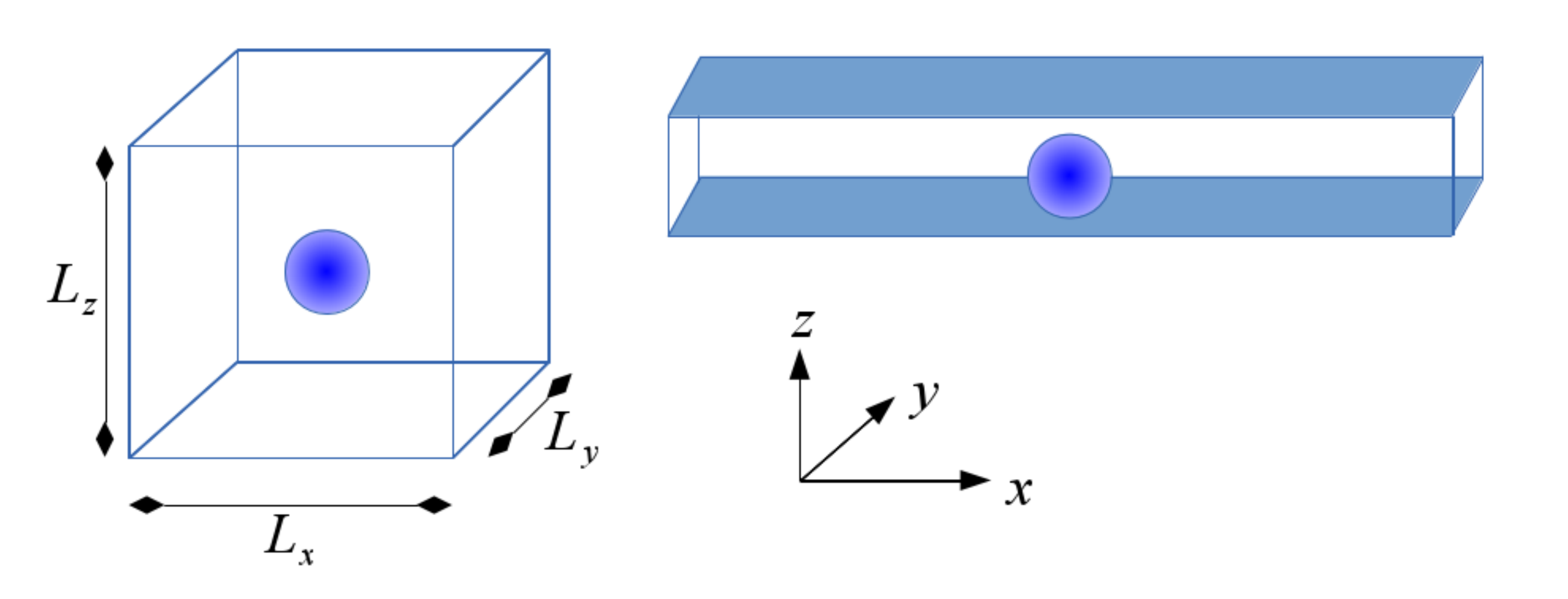}
\caption{Schematic picture of a spherical colloidal particle suspended in a binary solvent. 
$x$, $y$, and $z$ refer to the usual Cartesian coordinates and $L_x$, $L_y$, $L_z$ are the 
side lengths of the simulation box along these directions, respectively. Surfaces of the 
simulation box marked by white have periodic boundary condition and the blue colored surfaces 
bear symmetry breaking surface fields with preferential attraction to the same species of the 
solvent. In both cases, the colloid is placed at the centre of the box.}
\label{fig1}
\end{figure}

\cref{CHC6} and \cref{CHC7} have to be supplemented with appropriate boundary conditions (BC) 
\cite{diehl1992} at the colloid surface. One BC corresponds to no OP flux through the colloid 
surface
\begin{equation}\label{bc1} 
  ({\hat n} \cdot \nabla \mu(\vec r,t))|_{{\mathscr S}} = ({\hat n} \cdot \nabla\mathcal{F}[\psi]/\delta\psi(\vec r,t))|_{{\mathscr S}}=0,
\end{equation}
where, $\mathscr S$ and $\hat n$ refer to the colloid surface and the unit vector perpendicular 
to it and pointing into it, respectively. The other BC accounts for the adsorption preference of 
the colloid surface for one of the two components of the binary liquid mixture. This is done by 
considering a surface energy contribution 
$\frac{1}{2}\alpha \int_{\mathscr S} \psi^2 dS - h \int_{\mathscr S}\psi dS$ \cite{diehl1997} in 
addition to the bulk free energy $\mathcal F$. Here, $\alpha$ is a surface enrichment parameter 
and $h$ is the symmetry breaking surface field. After suitable rescaling of $\alpha$ and $h$, 
this gives rise to the dimensionless static so--called Robin BC
\begin{equation}\label{bc2} 
  ({\hat n} \cdot \nabla \psi(\vec r) + \alpha \psi(\vec r))|_{{\mathscr S}}=h.
\end{equation}
For the boundary condition associated with the temperature field we take
\begin{equation}\label{bc3} 
  \tilde T(\vec r)|_{\mathscr S} = {\tilde T}_1,
\end{equation}
which corresponds to maintaining the reduced temperature on the colloid surface at a constant 
value ${\tilde T}_1$; we assume no heat diffusion in the colloid. 

For the numerical setup the spherical colloid of radius $R$ is placed at the center of a 
rectangular box of side lengths $L_x,~L_y$ and $L_z$ (see \cref{fig1}). Outside the colloid, 
each grid point on a simple cubic lattice mesh refers to the binary solvent which is characterized 
by $\psi(\vec r,t)$ and $\tilde T(\vec r,t)$. The \textit{initial} configuration is generated as 
follows: each solvent grid point is assigned an OP value which is chosen from a uniform random 
number distribution $[-\frac{1}{2}: \frac{1}{2}]$ such that the spatially averaged OP is $\psi_0=0$. 
This corresponds to the critical concentration of the binary solvent. The \textit{initial} 
temperature values throughout the system are $\tilde T_i(\vec r)=1$. At $t=0$, the grid points 
corresponding to the colloid surface are quenched to a temperature $\tilde T_1$ and \cref{CHC6} 
and \cref{CHC7} are solved numerically using the Euler \cite{book-numerical} method to obtain 
$\psi(\vec r,t)$ and $\tilde T(\vec r,t)$. Periodic boundary conditions \cite{allen1987} are 
imposed at the side walls of the cubic box. In order to implement the BC on a curved colloid 
surface a trilinear interpolation method \cite{interpolation} has been used. All numerical 
results presented here correspond to a temperature quench inside the miscibility region of the 
equilibrium phase diagram $T_1=-1$ and they have been obtained using a numerical timestep 
$dt=0.001$. The spatially averaged order parameter is maintained at its critical value $\psi_0=0$. 
We consider homogeneous colloids constant value of surface parameters $\alpha$ and $h$. The 
noise amplitude is taken to be $10^{-4}$. Typical values of dimensional parameters used here are 
$a_0 = 0.2$ nm, $\xi_{-}(T_1) = 1.2$ nm, $\mathcal{A}\simeq 46 $, $t_0=10^{-4}$ s, 
$D_m = 10^{-13}$ m$^2$/s and $D_{th} = 10^{-7}$ m$^2$/s.

Note that by construction, \cref{CHC7} does not ensure that at the outer edge of the cubic 
simulation box the temperature is maintained at its initial value $\tilde T_i$ at all times. 
One of the possibilities is to keep the temperature at the boundary of the simulation box free 
with no heat flux through the outer edge of the simulation box:
\begin{equation}
\label{bcnf}
  \partial\tilde T(\vec r,t)/\partial r|_{\text{outer edge}} = 0.
\end{equation}
This may serve well to describe the early temperature evolution within small sample cells. 
In order to mimic experimental realizations in which a colloidal particle is placed in a 
macroscopically large cell, one can set the temperature at the outer edge of the simulation box
\begin{equation}\label{heatbc1} 
  \tilde T(\vec r,t)|_{\text{outer edge}}=\tilde T_i 
\end{equation}
In the present paper we consider both the free BC with no heat  flux through the boundary and 
the BC given by equation \cref{heatbc1}.

For the \textit{confined geometry}, the spherical colloid is placed at the center of a 
rectangular box with side length $L_z$ along the $z$-direction much smaller than the length along 
the other two directions: $L_z<< L_x$ and $L_x=L_y$. The outer edges of the box along the $x$ and 
$y$ directions (see \cref{fig1}) bear periodic boundary condition (PBC) \cite{allen1987} and the 
top and bottom edges along $z$ direction hold the surface BC
\begin{equation}\label{bc4} 
  ({\hat n} \cdot \nabla \psi(\vec r) + \alpha_s \psi(\vec r))|_{{\mathscr S}}=h_s.
\end{equation}
Both top and bottom surfaces prefer the same component of the binary solvent with the same strength. 
The top and bottom surfaces also hold the BC given by \cref{bc1}.

\section{Results}\label{results}

\subsection{Temperature profile}
\label{subsec:tp}

We have solved \cref{CHC7} with the BC on the surface of the colloid \cref{bc3} and the 
initial condition $\tilde T(\vec r,t=0)=\tilde T_i$ analytically using spherical coordinates 
for $R\leq r \leq L$. For the BC given by Eq.~(\ref{bcnf}), the stationary state is 
$ \tilde T^{fr}_s(r)=-|\tilde T_1|$. It is approached according to
\begin{equation}\label{sol1} 
  \tilde T(r,t) = -|\tilde T_1|+\frac{1}{r}\sum_{n=1}\mathcal{A}_n\sin\Big(\lambda_n(r-R)\Big)e^{-\mathcal D\lambda_n^2t},
\end{equation}
where $\lambda_n$ are positive solutions of the equation
\begin{equation}
 \tan\Big(\lambda_n(L-R)\Big)=\lambda_n L
\end{equation}
and 
the coefficient $\mathcal{A}_n$ is given by 
\begin{equation}
 \mathcal{A}_n=\frac{(\tilde T_i+|\tilde T_1|)\int_{r'=0}^{L-R}(r'+R)\sin \lambda_nr'dr'}{\int_{r'=0}^{L-R}\sin^2 \lambda_nr'dr'}.
\end{equation}

For the temperature at the outer boundary of the simulation box $r=L$ fixed at the value 
$\tilde T(r=L,t)$=$\tilde T_i$ we find
\begin{equation}\label{sol2} 
  \tilde T(r,t) = \tilde T_s^{fx}(r)+2R(\tilde T_i+|\tilde T_1|)\sum_{n=1}\frac{\sin \frac{n\pi}{L-R}(r-R)}{n\pi r} e^{-{\mathcal D}\frac{n^2\pi^2}{(L-R)^2}t},
\end{equation}
where the stationary profile $ \tilde T_s^{fx}(r) $ is given by 
\begin{equation}\label{sol2a} 
  \tilde T_s^{fx}(r) = \frac{L\tilde T_i+R|\tilde T_1|}{L-R}\frac{(r-R)}{r}  - |\tilde T_1|\frac{R}{r}. 
\end{equation}

\begin{figure}
\centering
\includegraphics*[width=0.4\textwidth]{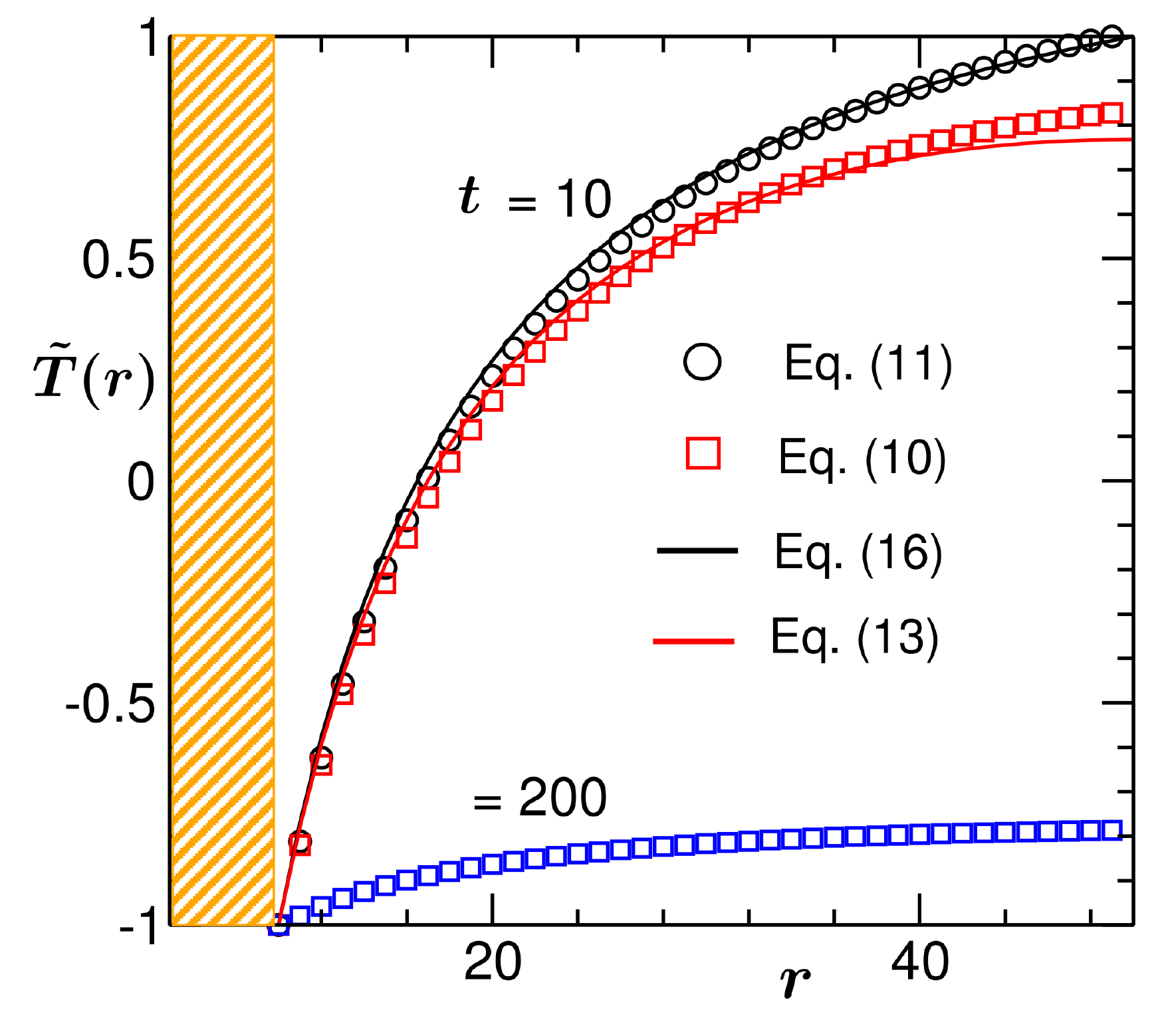}
\caption{Temperature profiles near a homogeneous colloid suspended in a binary solvent
at the time $t=10$ following a temperature quench of the colloid surface from $\tilde T_i=1$ 
to $\tilde T_1=-1$. The surrounding solvent cools due to heat flow from the colloid. 
$r$ is the radial distance in the midplane from the colloid center. Shaded region denotes 
the colloid surface. The square and circle symbols correspond to the numerical data for the 
temperature evolution following \cref{CHC7} with the BC given by \cref{bcnf} and by \cref{heatbc1}, 
respectively. In the case of BC given by \cref{heatbc1}, temperature of the solvent far away 
from the colloid is always maintained at the initial temperature $\tilde T_i=-1$, whereas using 
the BC given by \cref{bcnf} leads to lowering of temperature far away from the colloid. For the 
latter case, at $t=200$ the whole system is cooled down below the critical temperature and the 
temperature profile  is close to thesteady state one given by $\tilde T^{fr}_s(r)=-|\tilde T_1|$. 
The solid lines stand for corresponding analytical predictions for $\tilde T(r,t)$ (see \cref{sol1} 
and \cref{sol2}) with which the numerical data accord very well. For details of the numerical 
technique and analytical solutions, see main text. Results correspond to $L=100$, $R=10$, and 
$\mathcal D=50$. }
\label{fig2}
\end{figure}

In \cref{fig2} we compare the temperature profiles in the solvent evolving in time according 
to dynamical equation \cref{CHC7} with two different BC : (i) no heat flux at the outer edge of 
the simulation box  \cref{bcnf} and (ii) fixed temperature at the edge of the simulation box 
\cref{heatbc1}. There, $\tilde T(\vec r,t)$ is plotted against the reduced radial distance $r$ 
from the colloid centre in the midplane $z=L_z/2$ of the system at a fixed time $t=10$ (for no 
heat flux (i) BC we also show the temperature profile at the later time $t=200$ - see below). 
Results in \cref{fig2} correspond to $L_z=L_x=L_y=100$, $R=10$, $\tilde T_i=1$, $\tilde T_1=-1$. 
The shaded region is the colloid surface. For both types of BC the temperature at the surface of 
the colloid is maintained at the quench value $\tilde T_1=-1$ and with increasing distance from 
the colloid temperature gradually increases towards its initial value. However, in the case of 
fixed (ii) BC (circle), $\tilde T(\vec r,t)$  far away from the colloid is exactly equal to 
$\tilde T_i$. On the other hand, for no heat flux (i) BC (square) temperature away from the 
colloid is slightly lower than $\tilde T_i$. In \cref{fig2} we also plot the analytical expressions 
of $\tilde T(r,t)$ from \cref{sol1} and \cref{sol2}, marked by the solid lines with which our 
numerical data (symbol) for both agree very well. The slight discrepancy is due to the fact that 
the solution in \cref{sol2} is obtained within the spherical polar coordinate and the numerical 
results are obtained in a rectangular box and with a finite cubic grid. Note that a finer mesh 
will yield a much better matching between the two. In the case of no heat flux (i) BC the system 
continues to cool down until it reaches the stationary state with $ \tilde T^{fr}_s(r)=-|\tilde T_1|$. 
As can be inferred from the behavior of the temperature profile shown in \cref{fig2}, at time 
$t=200$ this stationary state is not yet achieved. 

\subsection{Coarsening in the bulk}
\label{subsec:cb}
\begin{figure}
\centering
\includegraphics*[width=0.45\textwidth]{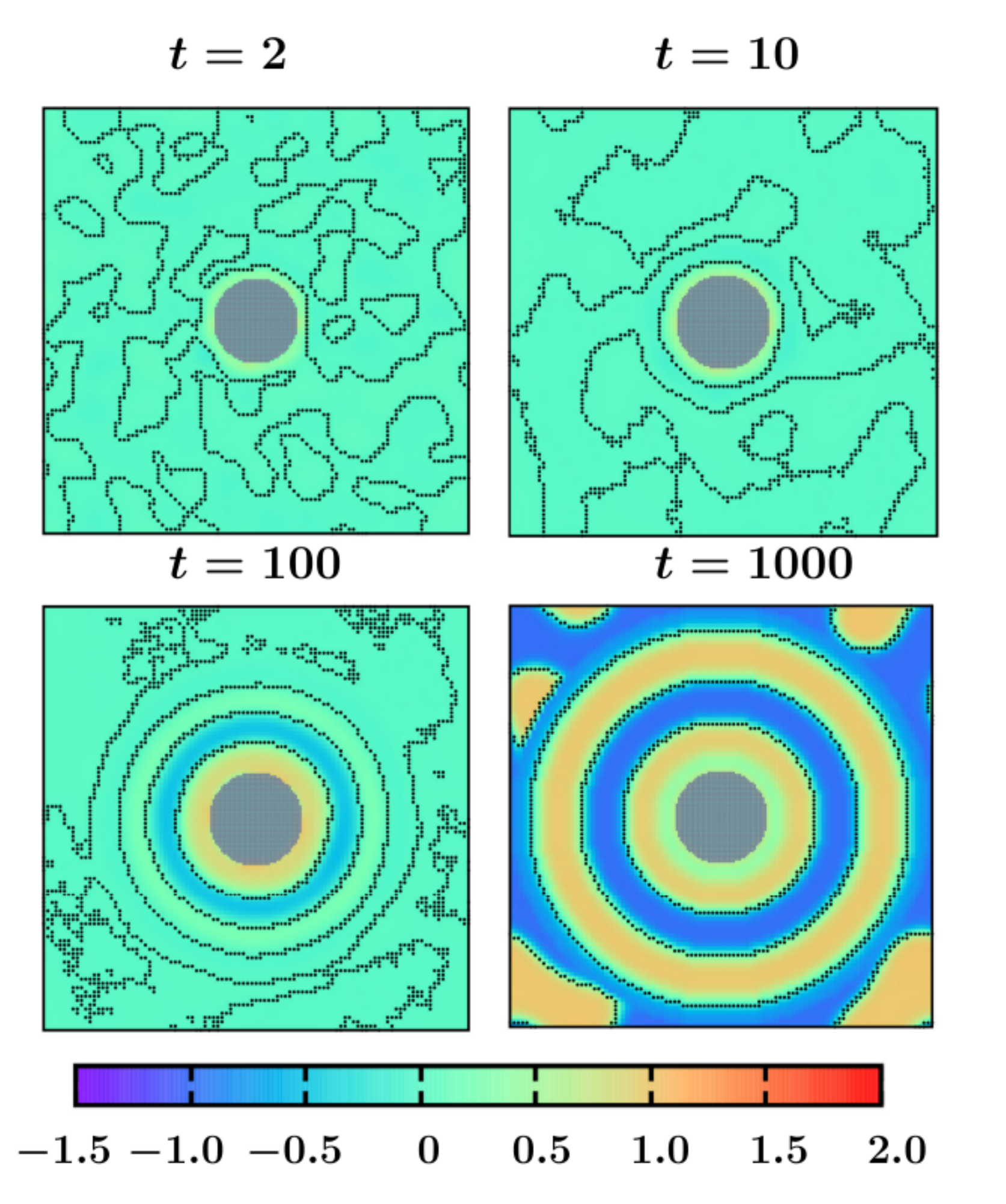}
\caption{Non-equilibrium coarsening of a binary solvent under time-dependent temperature 
gradient around the colloid with  preferential attraction to one of the two components of 
the binary mixture. Starting with a disordered initial configuration the colloid surface is 
cooled to $\tilde T_1=-1$ and the subsequent evolution of the order parameter and temperature 
fields are described by \cref{CHC6} and \cref{CHC7}, respectively. The BC for the temperature 
field at the outer edge of the simulation box is given by \cref{bcnf}. All snapshots are in 
the midplane $z=L_z/2$ of the system. The colorcode corresponds to different values of the order 
parameter with the black lines marking the contour lines and the grey circular region stands 
for the colloid. At early time, a spinodal-like decomposition in the bulk is present. With time, 
concentric circular surface rings form and they propagate into the bulk. Results corresponds to 
$L=100$, $R=10$, $\tilde T_1=-1$, $\alpha=0.5$, and $h=1$. }
\label{fig3}
\end{figure}

We first explore the effect of temperature evolution on the coarsening dynamics around the 
colloid in the bulk. Figure \ref{fig3} corresponds to no heat flux (i) BC. There we plot 
the coarsening snapshots at three different times following a temperature quench of the 
colloid surface from $\tilde T_0=1$ to $\tilde T_1=-1$. Results in \cref{fig3} correspond 
to a colloid with a homogeneous surface  with $\alpha=0.5$, and $h=1$, i.e.,  preferring 
a phase with $\psi >0$. The color code for the order parameter is provided in the figure, 
with the interface between the regions with $\psi >0$ and $\psi<0$ marked by black lines. 
Grey circle refers to the colloid. At very early time $t=2$ only one thin surface layer is 
formed on the colloid, whereas in the bulk spinodal-like phase segregation has started which 
is clear from the contour lines. With increasing time, these surface patterns propagate into 
the bulk and more concentric rings form. Note that two neighbouring rings consist of opposite 
phases. The qualitative feature of this process is similar to coarsening process under a 
temperature gradient for an off-critical concentration $\psi_0=0.1$ and with $\alpha=0.5$, 
and $h=1$ (see \cite{roy2017}). 

\begin{figure}
\centering
\includegraphics*[width=0.45\textwidth]{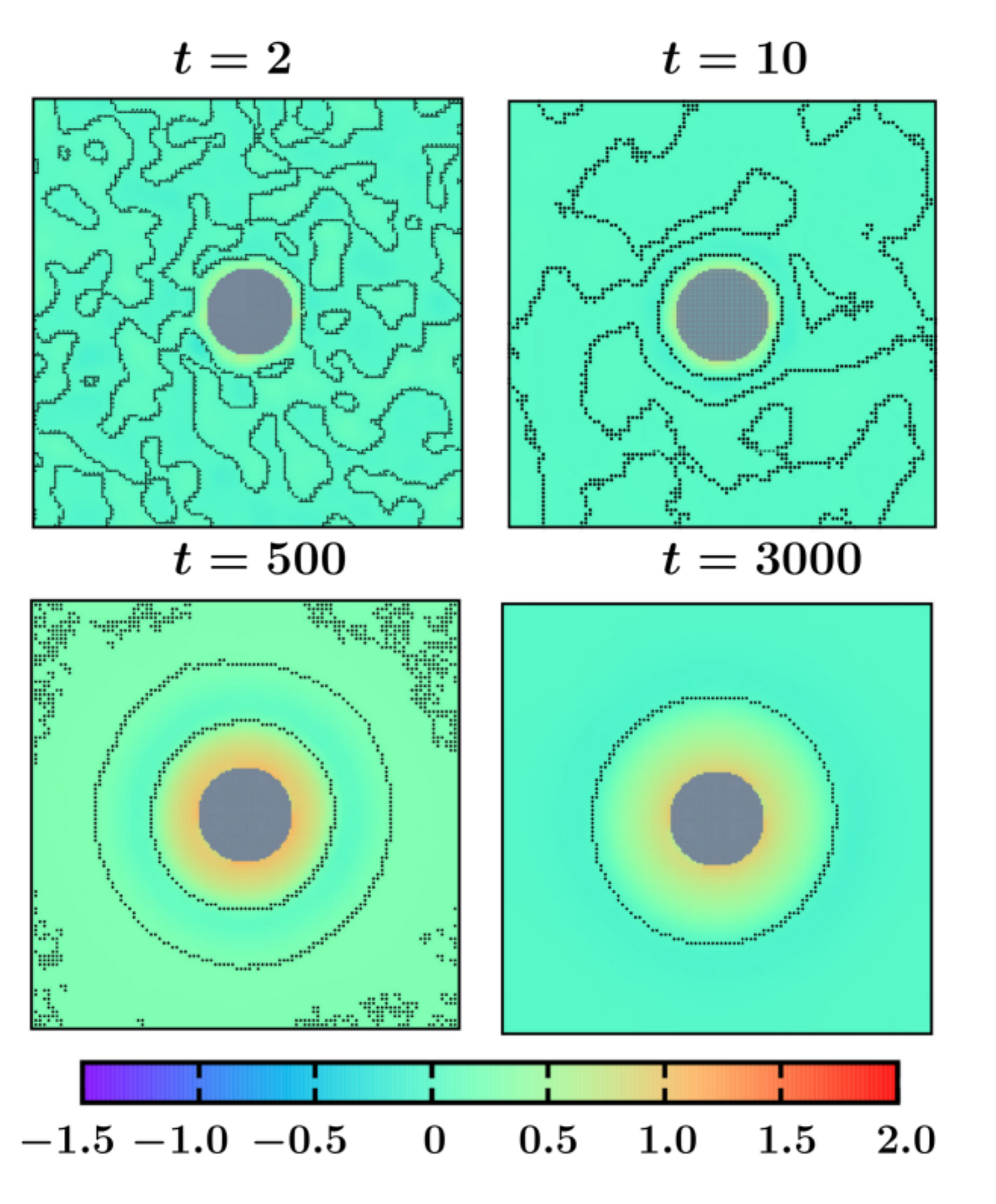}
\caption{Same as \cref{fig3}, but for the temperature evolution with the BC at the outer edge 
of the simulation box given by \cref{heatbc1}. In this case the ring-like surface patterns do 
not propagate much into the bulk and coarsening is much faster as compared to \cref{fig3}. }
\label{fig4}
\end{figure}

In \cref{fig4}, we present the snapshots of order parameter evolution for fixed (ii) BC. 
The surface patterns bear spherical symmetry like in \cref{fig3}. However, compared to \cref{fig3} 
the ring-like layers in this case propagate over a much shorter distance from the colloid. 
Also, coarsening progresses much faster in this case. Thus, different BCs for the temperature 
evolution (\cref{bcnf} and \cref{heatbc1}) do not change the qualitative features of the surface 
morphology, however, alter the distance over which the surface patterns propagate.

In order to further quantify the order parameter morphology shown in Figs.~\ref{fig3} and 
\ref{fig4}, we calculate the two-point equal time correlation function defined as 
\begin{equation}
 \label{eq:cor}
 C(\zeta=r-R,t)=\langle \psi(R,t) 
\psi(R+\zeta,t)\rangle-\langle \psi(R,t) \rangle \langle \psi(R+\zeta,t) \rangle
\end{equation}
in the midplane of the system. The symbol $\langle \cdot \rangle$ refers to the average over 
initial configurations of the angularly averaged $C(\zeta,t)$. 

\begin{figure}
\centering
\includegraphics*[width=0.4\textwidth]{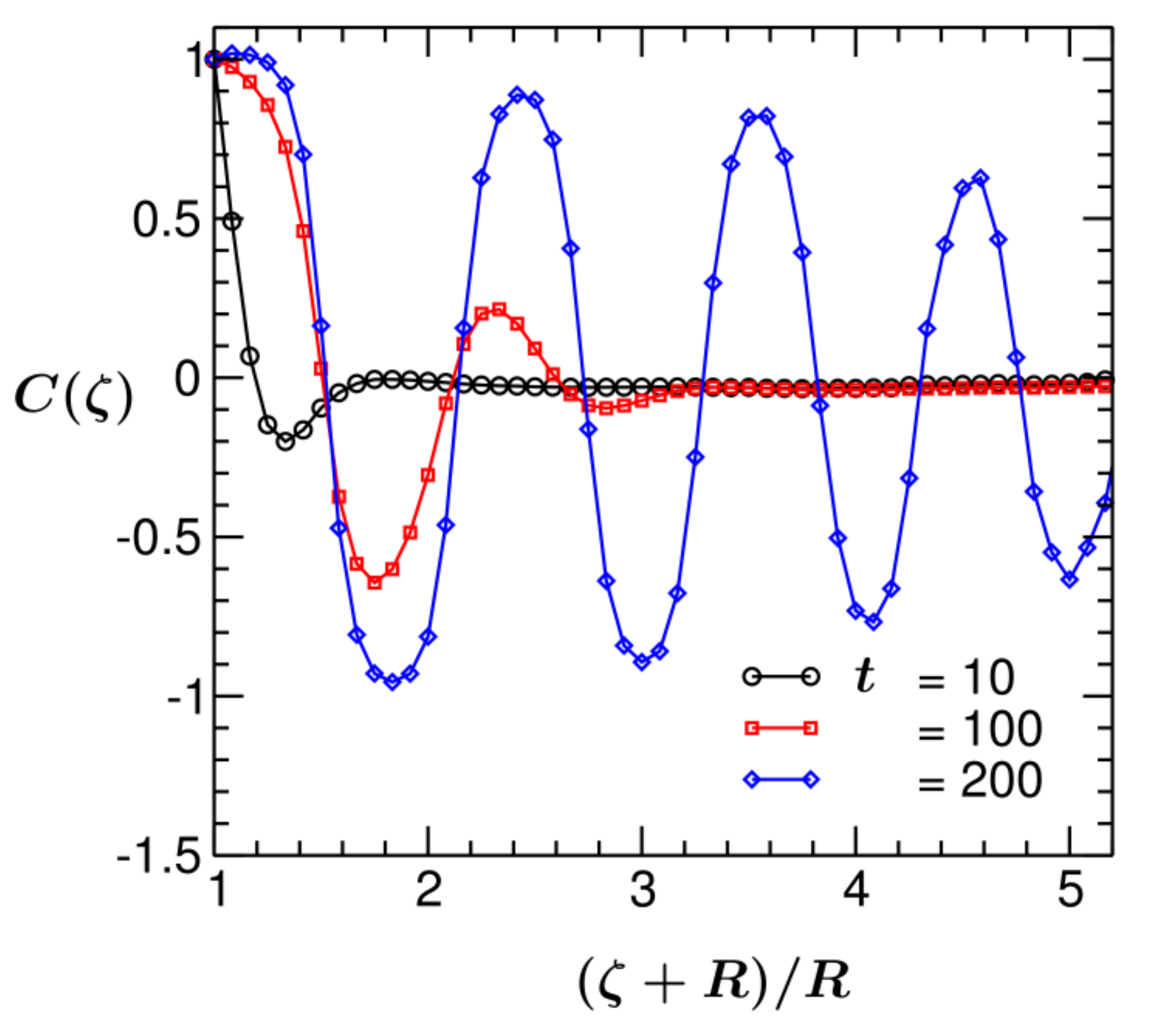}
\caption{Plot of the two-point equal time correlation function $C(\zeta=r-R)$ of the binary 
solvent vs. the scaled distance $(\zeta+R)/R$ in the midplane of the system, at three different 
times. Results correspond to the coarsening mechanism with a temperature gradient following 
\cref{CHC7} with no flux (i) BC \cref{bcnf} and for a colloid with preferential attraction 
to one of the two components of the binary solvent. Each maximum/minimum in $C(\zeta,t)$ 
corresponds to one surface layer around the colloid. With increasing time the thickness of 
the surface layer increases. All system parameters are same as in \cref{fig3}.}
\label{fig5}
\end{figure}

In \cref{fig5}, $C(\zeta,t)$ for the case of no flux (i) BC (corresponding to the snapshots 
shown in \cref{fig3}) is plotted vs. the rescaled distance $(\zeta+R)/R$, at three times $t$. 
At very early times, a spatial decay of $C(\zeta,t)$ is rather fast. Upon increasing time it 
slows down, the first layer widens and $C(\zeta,t)$ develops multiple peaks, which correspond 
to the various surface layers. As observed, at $t=10$  $C(\zeta,t)$ has one prominent minimum, 
indicative of one surface layer. However, at $t=200$  $C(\zeta,t)$ exhibits damped oscillations 
which extend to the boundary of the system. The corresponding temperature profile is shown in 
\cref{fig2}. Although at the time $t=200$ the temperature in the whole sample is  below the 
critical one, the pattern of phase separation is not that of a standard spinodal decomposition 
but remains a ring-like (see \cref{fig3}).

\begin{figure}
\centering
\includegraphics*[width=0.4\textwidth]{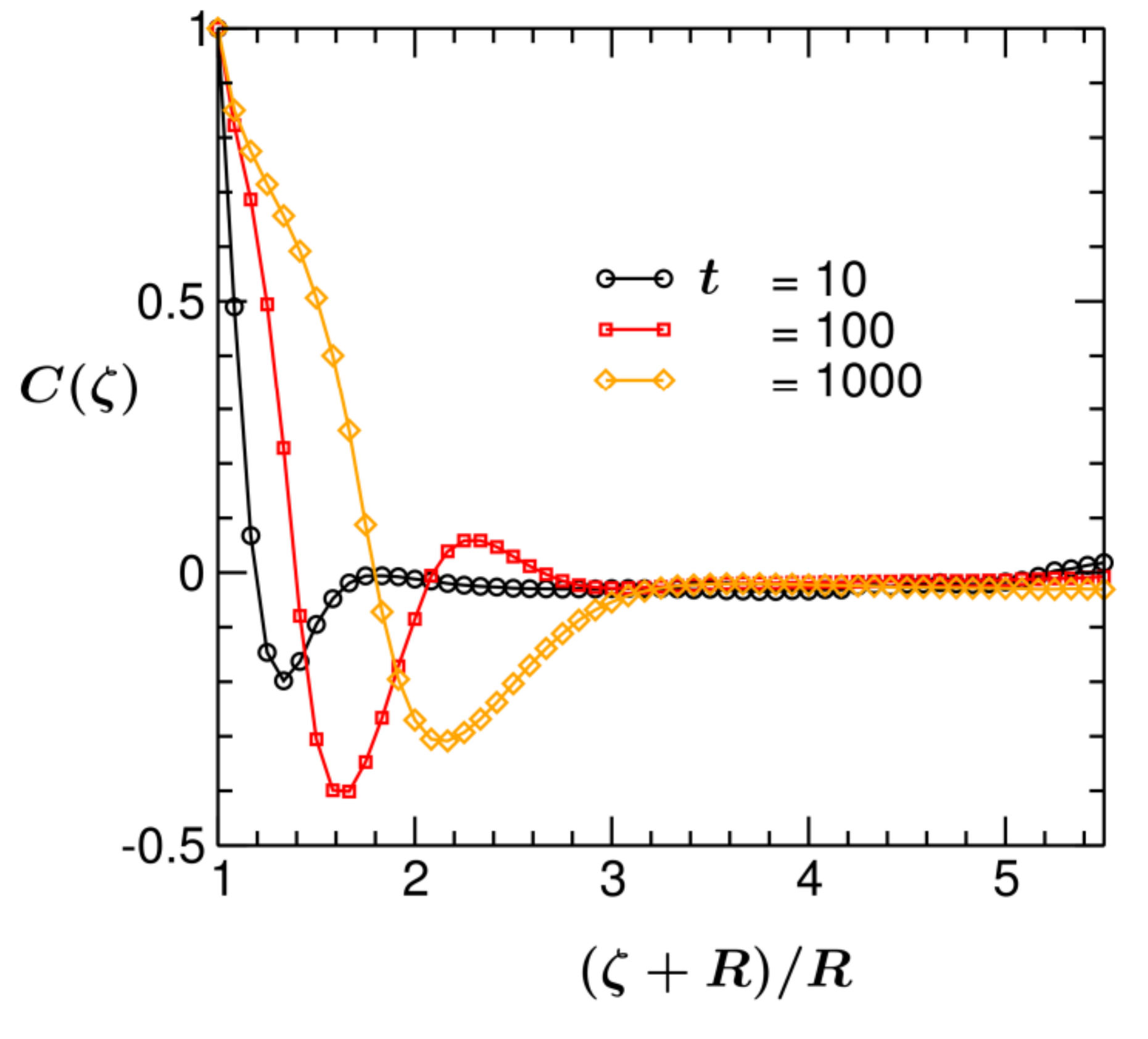}
\caption{The same as in \cref{fig5}, but for the temperature evolution followed by \cref{CHC7} 
with fixed (ii) BC \cref{heatbc1}. At very late times, $C(\zeta,t)$ exhibits only two mimima/maxima.}
\label{fig6}
\end{figure}

This has to be compared with the behavior of the correlation function for the case of fixed 
(ii) BC (corresponding to the snapshots shown in  \cref{fig4}). In \cref{fig6}, we plot 
$C(\zeta,t)$ vs. the scaled distance $(\zeta+R)/R$, at three different times. Qualitative 
trend of $C(\zeta,t)$ is similar to the one described above. However, one distinct feature 
is observed at very late times: for fixed (ii) BC  $C(\zeta,t)$ exhibits only two minima/maxima. 
This is due to the fact that for fixed (ii) BC the temperature further away from the colloid 
stays always above $T_c$ and the surface patterns can propagate over a much shorter distance 
from the colloid surface.

\begin{figure}
\centering
\includegraphics*[width=0.4\textwidth]{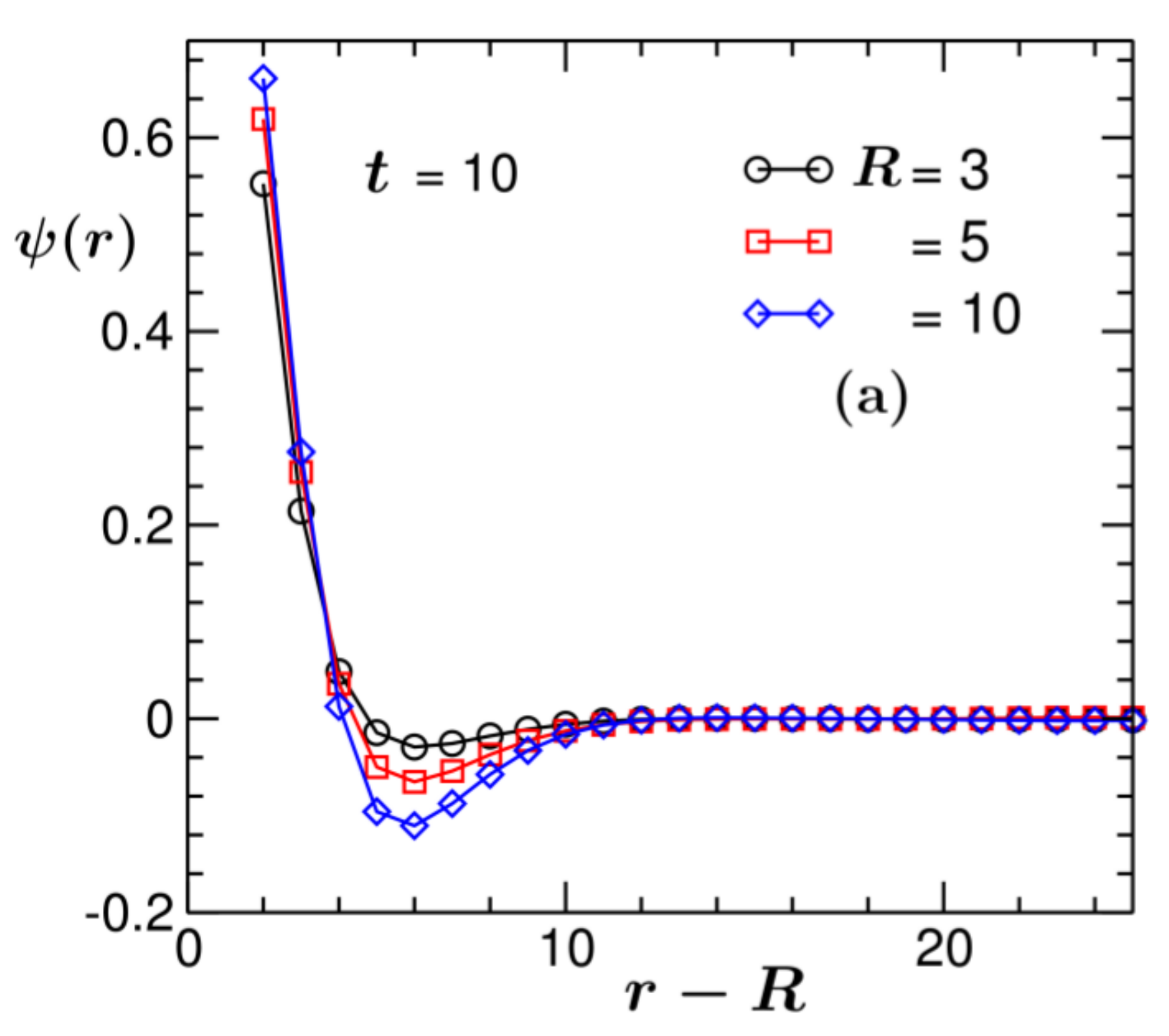}
\vskip 0.2cm
\includegraphics*[width=0.4\textwidth]{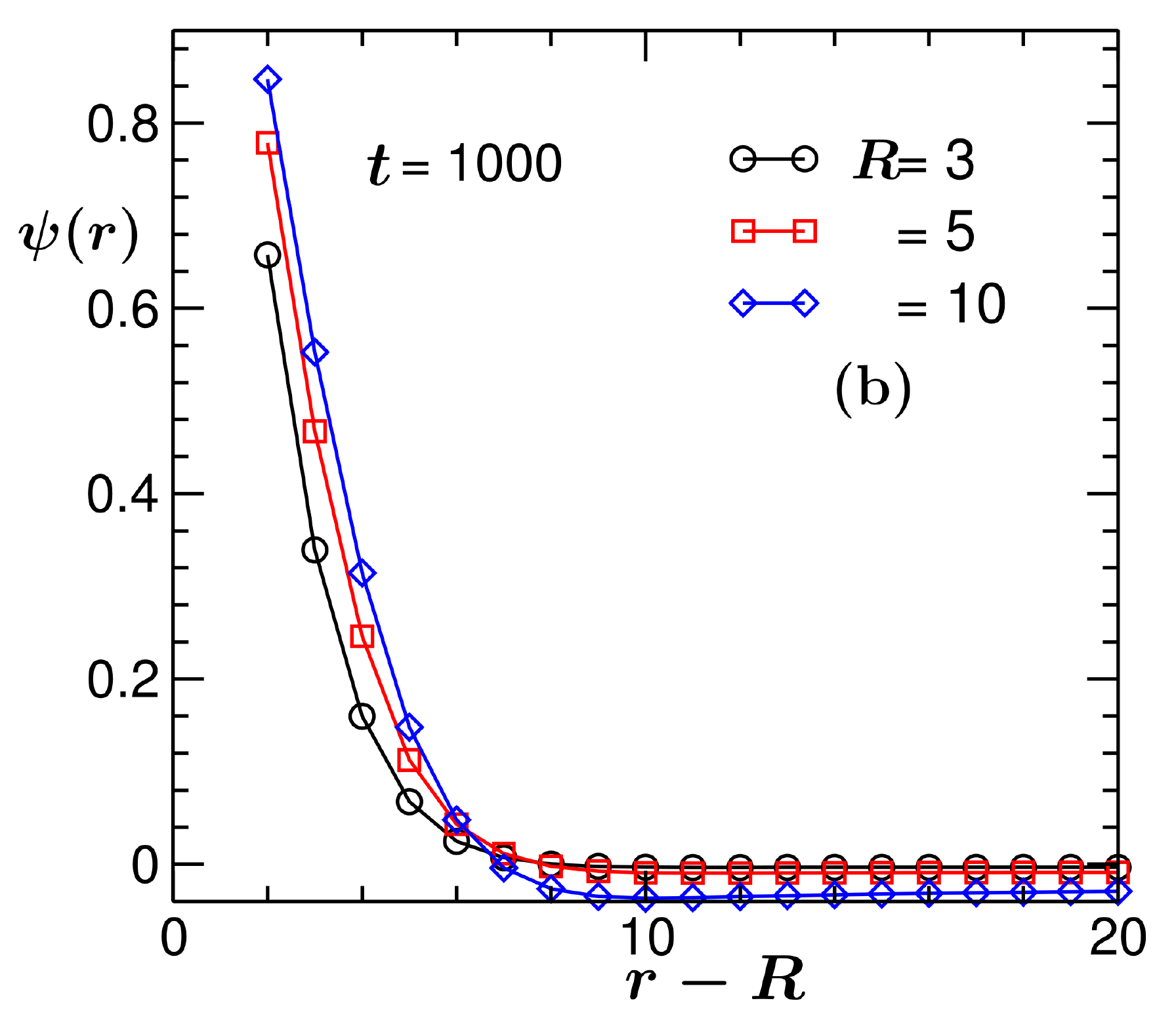}
\caption{Angularly averaged order parameter profile $\psi(\vec r,t)$ vs. the distance 
$(r-R)$ from the colloid surface at (a) an early time $t=10$ of a coarsening process 
and (b) at a very late time $t=1000$ when the system has almost reached the steady state. 
Different curves correspond to different colloid sizes. The temperature field evolves 
according to \cref{CHC7} with fixed (ii) BC \cref{heatbc1}. At both times, the value of 
the order parameter on the colloid surface increases with increasing radius $R$.}
\label{fig7}
\end{figure}

One factor that influences the early stage of the solvent coarsening is the size of 
colloidal particle. This can be inferred from \cref{fig7}(a), which shows the early 
time $t=10$ angularly averaged order parameter (OP) profile $\psi(\vec r,t)$ in the 
midplane $z=L_z/2$ of the system for (ii) BC. There, the results for three values of 
the colloid radius $R$ are presented. We find that upon roughly 3.3 fold increase of 
the colloid radius $R$,  the value of the OP on the colloid surface $\psi (0,t)$ 
increases monotonically by a factor of ca. 1.2. Concomitantly \textendash{} and in 
consistency with the conserved dynamics \textendash{} the minimum of $\psi(\vec r,t)$ 
deepens by approximately the same factor. Note that the position of the minimum remains 
unchanged. This is because, at least at early times, it is determined by the wave vector 
characterizing the fastest growing mode \cite{roy2017}. In \cref{fig7}(b), we also show 
the OP profiles at a late time $t=1000$ when the systems almost reached the steady state. 
In this case, with increasing $R$ by a factor of ca. 3.3, the value of the OP on the 
colloid surface increases by a factor of roughly 1.29.

\begin{figure}
\centering
\includegraphics*[width=0.45\textwidth]{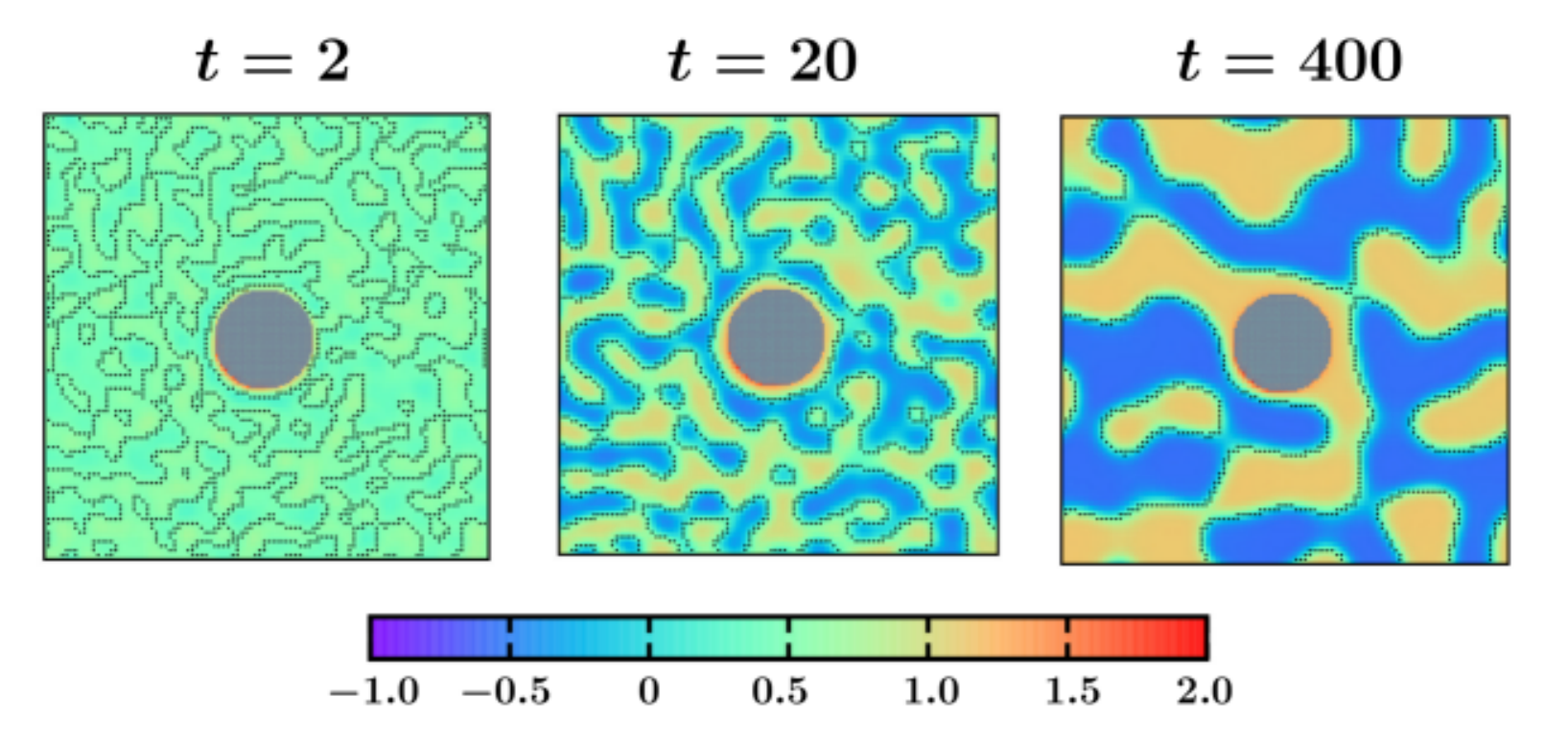}
\caption{Coarsening patterns around a colloidal particle for an instantaneous quench, 
i.e., when both the colloid and the solvent are at same constant temperature. Only one 
surface ring forms at early time. Spinodal-like patterns are more prominent, unlike in 
\cref{fig4}.}
\label{fig8a}
\end{figure}

It is instructive to compare the OP patterns evolving in the presence of a temperature 
gradient with those for an instantaneous quench, i.e., in the absence of any temperature 
gradient. The composition wave observed in the latter  case  (see \cref{fig8a}) is distinctly 
different from the one in \cref{fig4}. For the chosen values of $\alpha$ and $h$, only one 
`ring' forms and the spinodal-like patterns are more prevalent in the system. This observation 
is also in agreement with phase separation in polymer blends around fillers \cite{glotzer1999}. 
Of course, with time the thickness of the surface ring on the colloid surface increases. 
However, the qualitative and quantitative features of the coarsening patterns are different.

\begin{figure}
\centering
\includegraphics*[width=0.4\textwidth]{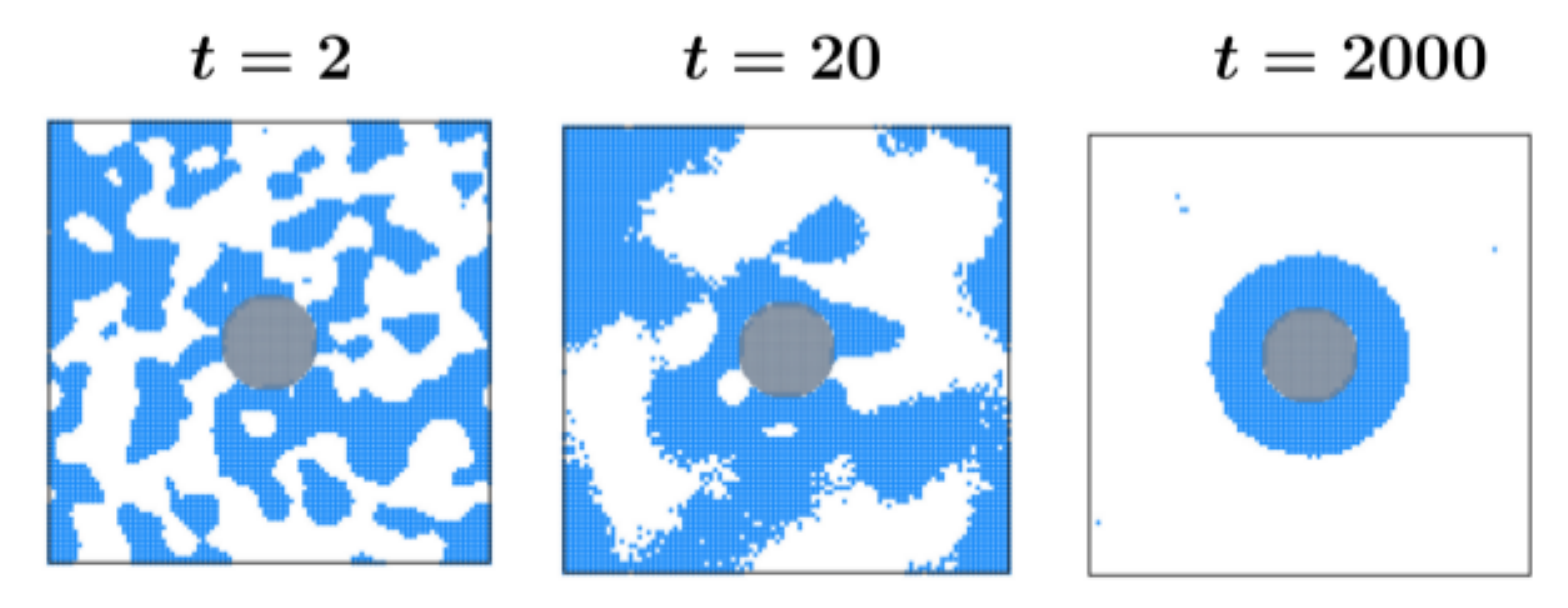}
\caption{Snapshots during the temperature-gradient induced coarsening of the binary 
solvent around a neutral colloid, i.e., without  preferential surface attraction for one 
of the two components of the binary solvent. Results correspond to $L=100$, $R=10$, $\tilde T=-1$, 
$\alpha=0$, $h_s=0$, and the midplane of the system. The temperature field obeys the 
flux free (i) BC. Blue and white parts stand for the two phases $\psi>0$ and $<0$, 
respectively. Starting from very early time both phases of the solvent are present 
on the surface of the colloid and the surface patterns do not bear spherical symmetry.}
\label{fig8b}
\end{figure}

Now, we focus on the influence of surface properties of the colloid on the coarsening 
morphology in the presence of a temperature gradient, for $\psi_0=0$. In \cref{fig8b}, 
we present the snapshots during the temperature-gradient induced coarsening around a 
neutral colloid, i.e., for $\alpha=0$ and $h=0$. The temperature field obeys the fixed 
temperature (ii) BC. Since $h=0$,  the order OP values are much smaller as compared to 
\cref{fig3} and \cref{fig4}. For the sake of clarity, in \cref{fig8b} we plot only the 
part of the solvent where $\psi(\vec r,t) >0$ in blue color. The white part corresponds 
to the other phase $\psi <0$. Clearly, starting from very early time the surface pattern 
is very different from the pattern for preferential attraction (see \cref{fig3} and \cref{fig4}). 
While in the case of preferential attraction with $\psi_0=0$ only one phase stays on the 
colloid surface at all times, for a neutral colloid both phases are present (see $t=2$ and 
$t=20$) until the system \textit{almost} completely phase separates at $t=2000$. Besides, 
the surface morphology for a neutral colloid is not spherically symmetric. 

In Ref.~\cite{roy2017} the solvent patterns around a neutral colloid during the coarsening 
process in the presence of temperature gradient have been studied for the off-critical 
concentration $\psi_0=0.1$. Comparing results from \cite{roy2017} with the present ones, 
we can see qualitative differences: for the critical concentration of the solvent, both 
phases form on the surface of a neutral colloid and the surface morphology is not 
spherically symmetric whereas for an off-critical concentration, only one phase is present 
on the surface of the neutral colloid and the surface patterns are spherically symmetric. 
Solvent concentration thus strongly affects the surface morphology for a neutral colloid. 

\subsection{Confinement effects}
\label{subsec:conf}
\begin{figure}
\centering
\includegraphics*[width=0.45\textwidth]{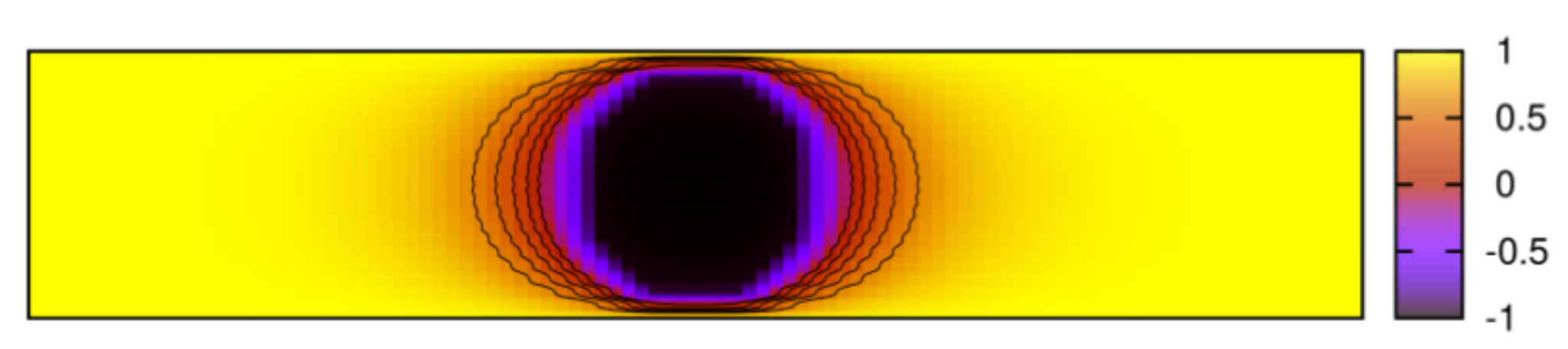}
\caption{Temperature profile around a colloidal particle confined in between two walls 
at $z=0$ and $z=L_z$ at time $t=10$  after a temperature quench from $\tilde T=1$ to 
$\tilde T_1=-1$. Both walls have preferential attraction to the same species of the binary 
solvent. Result corresponds to a vertical cut at $y=L_y/2$ in the $x-z$ plane and $L_x=100$, 
$L_z=20$, $R=8$, $\alpha=0.5$, and $h=1$. }
\label{fig9}
\end{figure}

In this section we investigate the effect of confinement on the non-equilibrium phase 
segregation dynamics around the colloid. This is highly relevant for experimental studies 
where typically  sample cells are finite slabs. \Cref{fig9} depicts the temperature profile 
of the binary solvent around the colloid at the time $t=10$ following the temperature quench 
from $\tilde T=1$ to $\tilde T_1=-1$ inside the binodal. Results correspond to the 
cross-section in the $x-z$ plane at $y=L_y/2$ with the confining walls at $z=0$ and $z=L_z$. 
Both walls and the colloid  prefer the same phase $\psi > 0$ of the binary solvent with 
the surface parameters $\alpha=0.5$ and $h=1$. A strong temperature gradient is observed 
in \cref{fig9}. The temperature at the outer boundaries of the simulation box are maintained 
at $\tilde T=1$ according to \cref{heatbc1} whereas close to the colloid $\tilde T(\vec r) < 0$. 
The spread of the region with negative temperature increases for larger colloids. 

\begin{figure}
\centering
\includegraphics*[width=0.35\textwidth]{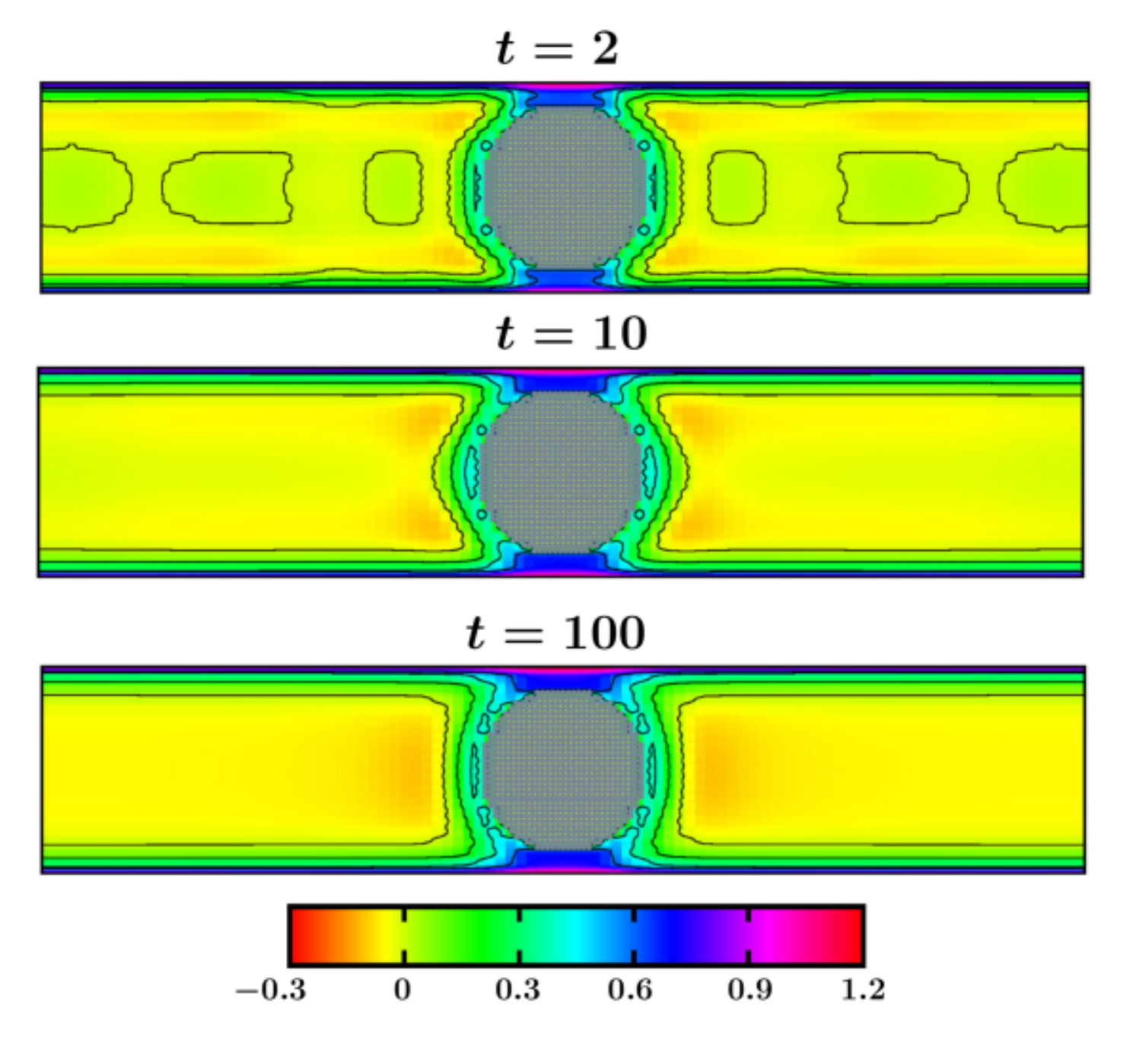}
\caption{Non-equilibrium order parameter evolution around a colloidal particle confined 
between two walls following a temperature quench from $\tilde T=1$ to $\tilde T_1=-1$. All 
parameters are the same as in \cref{fig9}. Results correspond to a thin slab with $L_z=20$ 
and $L_x=L_y=100$. A bridge connecting the colloidal particle and both the top and bottom walls
starts to form already at  very early time. With increasing time this bridge and the thickness 
of the surface layers grow and the spinodal-like phase separation proceeds in the bulk. }
\label{fig10}
\end{figure}

\begin{figure}
\centering
\includegraphics*[width=0.35\textwidth]{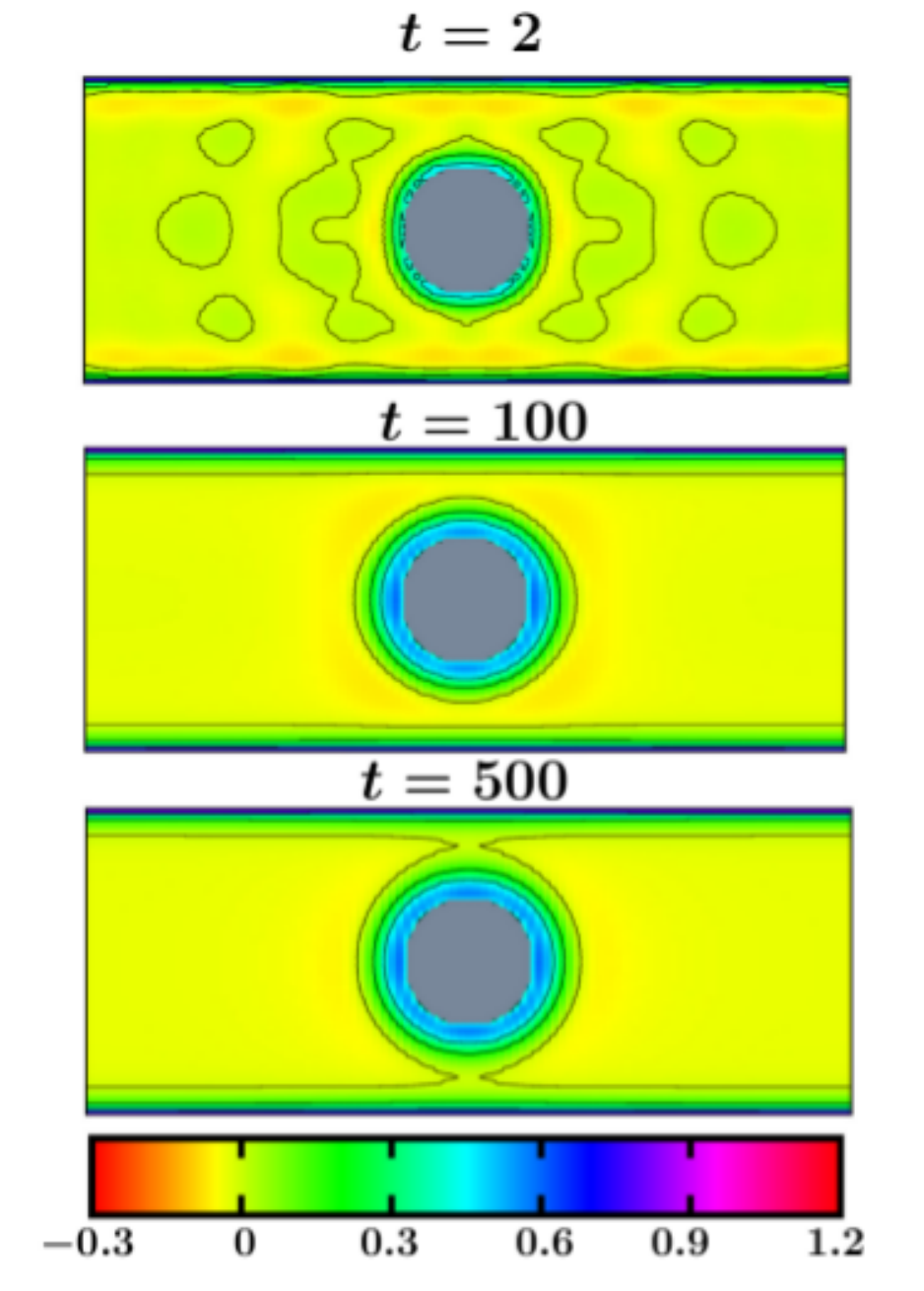}
\caption{Same as in \cref{fig10}, but for a thicker slab with $L_z=40$. At very early time 
the surface layers form around the colloidal particle as well as on the top and the bottom 
walls. As time increases, these surface layers get thicker and the spinodal-like phase 
separation proceeds in the bulk. Note that no prominent bridge formation is observed unlike 
in \cref{fig10}. }
\label{fig11}
\end{figure}

The evolution of the OP for a thin slab with $L_z=20$ and for a thick slab with $L_z=40$ is 
shown in \cref{fig10} and in \cref{fig11}, respectively. In both figures  $L_x=L_y=100$ and  
all other parameters are same as in \cref{fig9}. The color indicates the value of the OP. 
As can be seen in \cref{fig10}, already at very early time $t=2$ the surface layers form 
around the colloid as well as on the top and the bottom walls. These surface layers are all 
connected with each other, forming a bridge-like structure. Away from the confining surfaces 
exerting symmetry breaking BCs on the OP, the spinodal-like patterns are prominent. With 
increasing time the surface layers and the bridge get thicker and the bulk phase separation 
proceeds until a stationary state is achieved (see $t=100$, which is close to the stationary state). 
Note that the coarsening process in this case is much faster as compared to that in the bulk. 
This is because of the presence of the walls which attract one species of the solvent favorably 
and thus speed up  a phase  separation in the bulk. In the case of a thicker slab (see \cref{fig11}), 
at very early time $t=2$ the surface layers form on the colloid and the top and bottom walls, 
like in \cref{fig10}. However, in this case, the wall surface layers are separated from the 
one on the colloid and no bridge formation occurs. With increasing time the surface layers get 
thicker (see $t=100$) and the bulk phase separates. However, no thick bridge forms - even 
at very late time. Thus, coarsening of a solvent around the confined colloid depends strongly 
on the separation of the confining walls and this will have repercussion in multi-colloid 
dynamics as well.

\subsection{Soret term}
\label{subsec:st}
In the present  phenomenological approach, an effect of the inhomogeneous temperature field, 
the Ludwig-Soret effect, enters only through the OP flux due to the chemical potential gradient 
${\boldsymbol j} \propto - {\boldsymbol \nabla}\mu({\boldsymbol r},{t})$. However, following 
the Onsager's theory for irreversible processes, which provides phenomenological equations 
relating thermodynamic fluxes to the  generalized forces $-\mu/T$ and $1/T$ \cite{Onsager1931}, 
one can add to the OP flux ${\boldsymbol j}$ a cross-term  due to the temperature gradient with 
an independent coefficient related to the thermal diffusion coefficient $D_T$.  
\begin{equation}\label{soret1} 
  \frac{\partial \psi }{\partial t} = -\nabla \cdot (-M \nabla \mu) 
    -\nabla \cdot (-\frac{{\mathcal D_T}}{4}(\psi+1)(1-\psi)\nabla T).
\end{equation}
Here we assume that both $M$ and $\mathcal D_T$ are constants. Upon rescaling and adding the 
conserving noise \cref{soret1} reduces to
\begin{subequations}\label{soret2} 
\begin{align}
  \frac{\partial \psi(\vec r,t)}{\partial t}  &= \nabla ^2 \Big (\frac{{\tilde T}(\vec r,t)}{\tilde T_1} \psi(\vec r,t) + 
  \psi^3(\vec r,t) - \nabla^2 \psi(\vec r,t) \Big)\\ \nonumber
 & +  \frac{{\mathcal D}_T}{D_m}\frac{T_c}{4 {\mathcal A}}\nabla\Bigg( \Big( \psi_0^{-2}-\psi^2\Big)\nabla \frac{\tilde T(\vec r,t)}{|\tilde T_1|}\Bigg) 
  +\eta(\vec r,t)
  \end{align}
\end{subequations}
For typical binary solvent away from phase transitions the ratio ${\mathcal D}_T/{\mathcal D}_m$, 
which is called the Soret coefficient, is of the order of  $S_T \simeq 10^{-3} K^{-1}$. This means 
that for deep quenches the Ludwig-Soret contribution causes only a minor perturbation to the behavior 
of a system. Indeed, we have checked that for the values of the coefficient 
$B=({\mathcal D}_TT_c)/(4D_m{\mathcal A})$ ranging from 0 (i.e., no Soret effect) to 0.5 and with 
$\psi_0=0$ the changes in the OP profile and in the two-point equal time correlation function 
are minimal. Specifically, for the temperature field evolving according to  \cref{CHC7} with fixed 
(ii) BC (\cref{heatbc1}), the OP value $\psi(0,t=10)$ on the colloid surface reduces from 
0.7479 at $B=0$ to 0.7385 for $B=0.5$. The first minimum of both the OP and the normalized two-point 
equal time correlation function $C(r)$ also slightly decreases with increasing $B$; for the OP it 
takes up values -0.1862 for $B=0$ and -0.2004 at $B=0.5$ whereas for  $C(r)$ it reaches the value 
-0.4636 for $B=0$ and -0.5230 for $B=0.5$. We note that for the fixed temperature (ii) BC, the 
Soret term in the evolution equation \cref{soret2} for the OP breaks the conservation of the total OP. 
This is why the OP profile is slightly shifted towards negative values. This is not the case for 
no flux (i) BC. Nevertheless, also for the latter BC the overall changes in the OP and in
$C(r)$ in the studied range of parameter $B\leq 0.5$ are tiny.

The cross-term may become more important in the asymptotic critical regime close to $T_c$. 
(Here we refer to a recent review on the Soret effect \cite{Kohler2016}). This is because the 
interdiffusion constant $D_m$ shows the critical slowing down~\cite{HH1977}, i.e., the increase of 
characteristic diffusion time $\propto D_m$ upon approaching a critical point, whereas the 
thermal coefficient ${\mathcal D}_T$ does not. Specifically, in the asymptotic critical regime 
$D_m \propto |\tau|^{-\nu(1 +\eta_z)}$, where $\tau =(T-T_c)/T_c$, $\nu \simeq 0.63$ is the 
critical exponent of the bulk correlation length of the solvent, and $\eta_z\simeq 0.0063$ is the 
critical exponent of the solvent viscosity. Because $D_T \simeq const$, one has 
$S_T \propto |\tau|^{-\nu(1 +\eta_z)}$. This means that for very shallow quenches the Soret term 
\cref{soret2} can be even dominant.

\section{Summary and outlook}\label{summary}
In summary, using numerical simulations and analytical theory we have studied non-equilibrium 
early-time coarsening dynamics of a binary solvent around a suspended colloidal particle in a 
presence of the time-dependent temperature gradient. Following a sudden temperature quench of 
the colloid surface the surrounding solvent cools via heat diffusion. The ensuing order parameter 
field is solved using the modified Cahn-Hilliard-Cook equation which takes care of the coupling 
to a time-dependent temperature-field, in conjunction with the heat diffusion equation which 
dictates temporal evolution of the temperature field. The colloid surface attraction preference 
to one of the two components of the binary mixture is modeled by considering the symmetry-breaking 
surface field. Two types of boundary conditions for the temperature field -- mimicking different 
physical situations -- have been considered and their influence on the non-equilibrium dynamics 
have been explored in details. 

We have studied the coarsening process for different surface adsorption properties, concentration 
of the solvent and temperature evolution conditions. Under the time-dependent temperature gradient 
and for a colloid with selective surface adsorption, upon a temperature quench the spherical surface 
layers form close to the colloid, which with increasing time propagate into the bulk. Concomitantly, 
the thickness of the surface layer on the colloid increases with time. For a neutral colloid, i.e., 
without any surface adsorption preference, drastic changes in the coarsening morphology are observed. 
In the latter case, both phases prevail on the colloid surface whereas in the former case only one 
phase stays on the surface. Also, in the absence of preferential attraction the surface patterns 
are not spherically symmetric. These features of coarsening are reflected in the two-point equal time 
correlation function and the order parameter profiles. We have also provided a comparison with the 
coarsening phenomena occurring after an instantaneous quench, i.e., without any temperature gradient 
in the system. In the absence of the temperature gradient, the surface ring formation is much less 
pronounced as compared to the spinodal decomposition.

Our study also presents results on the coarsening process around a confined colloid in the 
presence of two confining surfaces with a preference to the same component of the binary mixture. 
In this case, enriched surface layers form both on the colloid as well as on the walls which make 
coarsening much faster in the confined geometry. For thin films a liquid bridge forms connecting 
the colloidal particle and the confining walls. These results are particularly important for 
experimental realizations where typically a colloidal suspension is confined in a quasi-two-dimensional 
chamber. It will be interesting to study Janus particles in such geometries and the role of 
bridging for the mechanism of self-propulsion.

We note that the details of pattern evolution such as, e.g., the extension of the surface patterns 
depend crucially on the BCs of the temperature field imposed at the outer edge of the system. 
The type of the BCs for the particular cases could be validated experimentally by comparing 
the coarsening morphologies. Qualitative features of our numerical results have been found in 
experiments with micron-sized colloids, which will be reported elsewhere. In future we will 
undertake an improvement of our phenomenological model to include heat flow across the colloidal 
particle, which will better mimic experimental situations, and to account for heat dissipation.

\textbf{Acknowledgments:} The work by AM has been supported by the Polish National Science Center 
(Harmonia Grant No. 2015/18/M/ST3/00403). We thank Mihail N. Popescu for an inspiring discussion.


\begin{thebibliography}{100}
\bibitem{AY} R. Akbarzadeh, and A.M. Yousefi,  Journal of Biomedical Materials Research  Part B - Applied Biomaterials {\bf 102}, 1304 (2014).
\bibitem{Torino:2016} E. Torino, R. Aruta, T. Sibillano, C. Giannini, and P. A. Netti, Scientific Reports {\bf 6}, 32727 (2016).
\bibitem{binder2006} S. K. Das, S. Puri, J. Horbach, and K. Binder, Phys. Rev. Lett. \textbf{96}, 016107 (2006).
\bibitem{puri2005} S. Puri, J. Phys.: Condens. Matter {\bf 17}, R1 (2005).
\bibitem{BPDH} K. Binder, S. Puri, S. K. Das, and J. H\"orbach, J Stat Phys  {\textbf 138}, 51–84 (2010).
\bibitem{Chung} H.-J. Chung, K. Ohno, T. Fukuda, and R. J. Composto, Nano
Lett. {\bf 5}, 1878 (2005).
\bibitem{Herzig} E. M. Herzig, K. A. White, A. B. Schofield, W. C. K. Poon, and
P. S. Clegg, Nat. Mater. {\bf 6}, 966 (2007).
\bibitem{Stratford} K. Stratford, R. Adhikari, I. Pagonabarraga, J.-C. Desplat, and
M. E. Cates, Science {\bf 309}, 2198 (2005).
\bibitem{Krekhov:2013} A. Krekhov, V. Weith, and W. Zimmermann, Phys. Rev. E {\bf 88} 040302(R) (2013).
\bibitem{Iwashita:2013} Y. Iwashita and Y. Kimura,  Soft Matter {\bf 9} 10694 (2013).
\bibitem{Kurita:2017} R. Kurita, Scien. Rep. {\bf 7}, 6912 (2017).
\bibitem{Koehler} W. K\"ohler, A. Krekhov, and W. Zimmermann, Adv. Polym. Sci. {\bf 227}, 145 (2007).
\bibitem{Sano} H-R Jiang, N. Yoshinaga, and M. Sano, Phys. Rev. Lett. \textbf{105}, 268302 (2010).
\bibitem{wurger2013} T. Bickel, A. Majee, and A. W\"{u}rger, Phys. Rev. E. \textbf{88}, 012301 (2013).
\bibitem{bechinger2011} G. Volpe, I. Buttinoni, D. Vogt, H-J. Kammerer, and C. Bechinger
Soft Matter {\bf 7}, 8810  (2011).
\bibitem{buttinoni2012} I. Buttinoni, G. Volpe, F. K\"ummel, G. Volpe, and C. Bechinger, J. Phys.: Cond. Mat. {\bf 24}, 284129 (2012).
\bibitem{ruben2016} J.R. Gomez-Solano, A. Blokhuis, and C. Bechinger, Phys. Rev. Lett. \textbf{116}, 138301 (2016).
\bibitem{solano2017} J.R. Gomez-Solano, S. Samin, C. Lozano, P. Ruedas-Batuecas, R. van Roij, C. Bechinger, Scien. Rep. {\bf 7}, 14891 (2017). 
\bibitem{roy2017} S. Roy, S. Dietrich, and A. Maciolek, Phys. Rev. E {\bf 97},  042603  (2018).
\bibitem{Hohenberg1977} P. C. Hohenberg, B. I. Halperin, Rev. Mod. Phys. {\bf 49}, 435 (1977).
\bibitem{essery1990} R.C. Ball and R.L.H. Essery, J. Phys.: Condens. Matter \textbf{2}, 10303 (1990).
\bibitem{binder2013} P. K. Jaiswal, S. Puri, and K. Binder, EPL {\bf 103}, 66003 (2013).
\bibitem{Kawasaki} K. Kawasaki, {\it Phase Transition and Critical Phenomena}, edited by C. Domb and 
M. S. Green, V. {\bf 2} (Academic Press, London) 1972.
\bibitem{diehl1992} H. W. Diehl, and H. K. Janssen,  Phys. Rev. A \textbf{45}, 7145 (1992).
\bibitem{diehl1997} H. W. Diehl, Int. J. Mod. Phys. B \textbf{11}, 3503 (1997).
\bibitem{book-numerical}J.C. Butcher, \textit{Numerical methods for ordinary differential equations} 
(Willey, England) 2008.
\bibitem{allen1987} M.P. Allen and D.J. Tildesley, \textit{Computer Simulations of Liquids}
(Clarendon, Oxford, 1987).
\bibitem{interpolation} L.M. Surhone, M.T. Timpledon, and S.F. Marseken, \textit{Trilinear Interpolation} 
(Betascript Publishing, United States, 2010).
\bibitem{glotzer1999} B. P. Lee, J. F. Douglas, and S. C. Glotzer, Phys. Rev. E {\bf 60}, 5812 (1999).
\bibitem{Onsager1931}  L. Onsager, Physical Review {\bf 37}, 405 (1931).
(1931). 
\bibitem{Kohler2016}  W. K\"ohler and  K. Morozov, J. Nonequilib. Thermodyn. {\bf 41}, 151 (2016).
\bibitem{HH1977}P. C. Hohenberg and B. I. Halperin BI  Rev Mod Phys {\bf 49}, 435 (1977).


\end{thebibliography}
\end{document}